# Defining and Computing Alternative Routes in Road Networks

Jonathan Dees, Robert Geisberger, and Peter Sanders

Karlsruhe Institute of Technology (KIT), 76128 Karlsruhe, Germany

jonathan.dees@student.kit.edu, {geisberger,sanders}@kit.edu

Roland Bader

BMW Group Research and Technology, 80992 Munich, Germany, roland.bader@bmw.de

October 23, 2018

## Abstract

Every human likes choices. But today's fast route planning algorithms usually compute just a single route between source and target. There are beginnings to compute *alternative routes*, but this topic has not been studied thoroughly. Often, the aspect of *meaningful* alternative routes is neglected from a human point of view. We fill in this gap by suggesting mathematical definitions for such routes. As a second contribution we propose heuristics to compute them, as this is NP-hard in general.

# 1 Introduction

The problem of finding the shortest path between two nodes in a directed graph has been intensively studied and there exist several methods to solve it, e.g. Dijkstra's algorithm [Dij59]. In this work, we focus on graphs of road networks and are interested not only in finding *one* route from start to end but to find *several* good alternatives. Often, there exist several noticeably different paths from start to end which are almost optimal with respect to length (travel time). For a human, it can have advantages to be able to choose the route for his tour himself among a set of good alternatives. A person may have personal preferences or knowledge for some routes which are unknown or difficult to obtain, e.g. a lot of potholes. Also, routes can vary in different attributes beside travel time, for example in toll pricing, scenic value, fuel consumption or risk of traffic jams. The trade-off between those attributes depends on the person and the persons situation and is difficult to determine. By computing a set of good alternatives, the person himself can choose the route which is best for his needs.

There are many ways to compute alternative routes, but often with a very different quality. In this work, we propose new ways to measure the quality of a solution of alternative routes by mathematical definitions based on the graph structure. Also, we present several different heuristics for computing alternative routes as determining an optimal solution is NP-hard in general.

## 1.1 Related Work

Computing the $k$-shortest paths [Epp94, Yen71] computes the $k$ shortest paths as alternative routes and regards sup-optimal paths. The computation of disjoint paths is similar, except that the paths must not overlap. [Sco97] proposes a combination of both methods: The computation of a shortest path, that has at most $r$ edges in common with the shortest path.

Another approach uses several edge weights to compute Pareto-optimal paths [Han79, Mar84, DW09]. Given a set of weights, a path is called Pareto-optimal if it is better than any other paths for respectively at least one criteria. All Pareto-optimal paths can be computed by a generalized Dijkstra's algorithm.

The *penalty* method iteratively computes shortest paths in the graph while increasing certain edge weights [CBB07]. [SS07] present a speedup technique for shortest path computation including edge weight changes.

Alternatives based on two shortest paths over a single *via node* are considered by the Plateau method [CAM09]. It identifies fast highways (plateaus) which define a fastest route from $s$ to $t$ via the highway (plateau).

# 2 Alternative Graphs

Our overall goal is to compute a set of alternative routes. However, in general, they can share nodes and edges, and subpaths of them can be combined to new alternative routes. So we propose the general definition of an *alternative graph* that is the union of several paths from source to target. More formally, let $G = (V, E)$ be a graph with edge weight function $w : E \to \mathbb{R}_+$. For a given source node $s$ and target node $t$ an alternative graph $H = (V', E')$ is a graph such that for every edge $e \in E'$ there exist an $s$-$t$-path in $H$ containing $e$ and for every node $u \in V' \subseteq V$ there exists an $s$-$t$-path containing $u$. Furthermore, every edge $(u, v)$ in $E'$ must represent a path from $u$ to $v$ in $G$ with the same weight as the edge.

We call an alternative graph *reduced* iff we contract all nodes in $V' \setminus \{s, t\}$ that have only one incoming and one outgoing edge. A reduced alternative graph provides a compact representation of all alternatives as it only contains those nodes where either several alternative paths split up or unite.

We focus here on the computation of an alternative graph. The extraction of several alternative routes from the alternative graph depicts another problem as we cannot just enumerate all paths in this graph (e.g. when there are loops). As the alternative graph is usually very small, we can use complex algorithms, including ranking and user preferences, to extract routes.

# 3 Attributes to Measure in Alternative Graphs

Our goal is to find meaningful alternative graphs from a human perspective. Still we need to quantify the quality of alternative graphs. So we propose several attributes to measure. The combination of those attributes enables us to evaluate alternative graphs.

Given an alternative graph $H = (V', E')$, we introduce an auxiliary function pos : $V' \to [0..1]$ returning the relative position of a node $u \in V'$ on a shortest $s$-$t$-path in $H$ via $u$. It is defined as $\text{pos}(u) := \frac{d_H(s,u)}{d_H(s,u)+d_H(u,t)}$ where $d_H(u,v)$ returns the shortest distance from $u$ to $v$ in $H$.

One thing we want to measure for an alternative graph, is how many alternatives are included in the graph. The first idea is to count the number of different paths from $s$ to $t$. However, this hardly reflects the amount of choice. For example, the number of different paths increases exponential for short branches in a row but is rather low for large parallel branches, which is counterintuitive. Instead, we measure the amount of choice by the (normed) total distance in the graph. Having two totally disjoint paths from $s$ to $t$, we have a total distance of 2, independent of the length of the paths. So we do not want that long paths artificially increase the amount of choice. This leads to the following definition for total distance in alternative Graph $H = (V', E')$:

$$\text{total distance} = \sum_{e=(u,v) \in E'} \frac{w(e)}{d_H(s,u) + w(e) + d_H(v,t)}$$

The lengths of the alternatives are important, they should be close to the optimum. Using the definition of total distance, we define the (normed) average distance. This reflects the average length of alternatives in the alternative graph $H = (V', E')$ by multiples of the length of the shortest route from $s$ to $t$ in the base graph $G = (V, E)$:

$$\text{average distance} = \frac{\sum_{e \in E'} w(e)}{d_G(s,t) \cdot \text{total distance}}$$

Every node in the alternative graph having more than one outgoing edge implies a decision, which outgoing edge to take. The number of those split ups is important, as a human cannot handle too many of them. The number of decision edges is equal to the number of edges minus one in the reduced alternative graph.

$$\text{decision edges} = \sum_{v \in V' \setminus \{t\}} \text{outdegree}(v) - 1$$

Considering an alternative graph with three disjoint alternative paths at the first half and an alternative graph with two disjoint alternatives each at the first and the second half, the latter case is more attractive as the distribution is better. In the optimal case, we have the same amount of alternative paths at each position in the alternative graph. For comparison, we measure the variance of the distribution of the paths. The average number of edges at a position is equal to the total distance and the function number_of_edges_at_pos(x) returns the number of edges $(u,v)$ in the alternative graph, for which $\text{pos}(u) \leq x \leq \text{pos}(v)$.

$$\text{variance} = \int_0^1 (\text{total distance} - \text{number\_of\_edges\_at\_pos(x)})^2 \text{dx})$$

The variance is varying more if the total distance is high, for normalization, we compute the coefficient of the variation.

$$\text{coefficient of variation} = \frac{\sqrt{\text{variance}}}{\text{average number of edges}}$$

The above attributes are just a few of a plethora of possibilities to measure. Further attributes to measure are e.g. the enclosed area of the alternative graph considering the geographical positions from nodes and edges or the number of possible paths (without loops) in the alternative graph.

# 4 Combining Measurements as Objective Function for Optimization

We want to quantify the quality of alternative graphs using the defined attributes from the last section. Observing only one attribute from above is not meaningful as there exist alternative graphs with very good values for a single attribute while being a degenerated graph, i.e. without useful alternatives. That means, we must observe several attributes at a time. To compare the quality of two alternative graphs, we need to compare several attributes. If one alternative graph dominates every other attribute of the other, the comparison is easy. However, if no alternative graph dominates the other with respect to the attributes, it is difficult to identify the better solution. Our goal is to quantify the quality in one single number. Therefore, we combine the attributes to a single number (target function) while using some constraints on the attributes.

In case of the proposed attributes, a good alternative graph

- maximizes the total distance.

- minimizes the average distance.

- minimizes the decision edges (On the first view this seems to be counterintuitive as it limits the possibilities for alternatives, however, minimizing the decision edges improves the quality of single branches when maximizing total distance and minimizing average distance. As only a certain amount of decision edges is useful for a human, this is also good attribute for a constraint.).

- maximize the size of the enclosed area (this favors to routes with higher geographical distance to each other, which are less correlated for traffic jams).

- maximize the number of different paths (without loops) in the alternative graph.

If we combine the attributes to a single target function with usefull constraints, computing the optimum is likely to be NP-hard. For computing good alternative graphs, we present several heuristics in the next section.

# 5 Methods to Compute Alternatives

A meaningful combination of measurements is most likely NP-hard to optimize. Therefore, we restrict ourselves to heuristics to compute an alternative graph. We present several known methods and some new ones, and plan to evaluate them based on the attributes of Section 3.

## 5.1 Pareto

A classical approach to compute alternatives is Pareto optimality. As we cannot use the measures for alternative graphs during the Pareto path computation, we need to use something else. A possibility is to use several different metrics (travel time, distance, scenic value, etc.) or to combine it with e.g. a single-criteria query and mark the edges on the shortest paths with 1 and all other with 0 (or their percentage distance on the shortest path). Or we take the distance to the shortest path into account and prefer routes that are farther away. Therefore we get for any percentage of deviation from the shortest path the best possible alternative. Still, this does not perform well. First of all, there are too many alternatives and especially there are a lot with only small deviations. Still, there are some interesting alternatives. Therefore, we tighten the domination criteria to keep only paths that are sufficiently different. We suggest two methods for tightening described in [DW09]. First, we can reduce the computed paths by saying that, for a given $\epsilon > 0$, path (*or subpath*) $p_1$ dominates path (or subpath) $p_2$, if $p_2$ is at least $\epsilon$ times longer, in addition to the regular domination. That makes sense, as we are not interested in paths which length is far from optimum. Additionally, we can further tighten the dominance, by saying that a path (*or subpath*) $p_1$ also dominates path $p_2$, if $p_2$ is longer but not essentially better in the other metrics. More formally, $p_1$ with attributes (length$'$,$w_1$,$w_2$,...) dominates $p_2$ with attributes (length$'$,$w'_1$,$w'_2$,...) if $\frac{\sum w'_i}{\sum w_i} > \frac{\text{length}}{\gamma \cdot \text{length}'}$ for some constant $\gamma$. This allows longer paths only if they are enough different from the shortest path. Besides filtering unwanted paths, it significantly improves the computation time and reduces the memory footprint. Both methods reduce the set of found paths by arguably mostly bad alternative routes. Unfortunately, we cannot guarantee that all good alternative routes are Pareto optimal. However, relaxing domination criteria may be difficult as it would again result in too many unwanted alternatives. Even when tightening the domination criteria, it is crucial which computed paths are selected to build up our alternative graph. Therefore, we iteratively select a path among all computed ones in a greedy fashion that maximises our target function of Section 4 while still satisfying the constraints.

## 5.2 $k$-Shortest Paths

A widely used approach [Epp94] is to compute the $k$ shortest paths between $s$ and $t$. So also suboptimal paths are regarded. However, these routes are very similar to each other, and only small changes are not relevant for humans. Computing all shortest paths up to a number $k$ produces many paths that are almost equal and do not "look good". So we need to identify and filter the best routes, similar to the Pareto method, but here we have a lot more similar paths. If we compute a very large number of shortest paths, we just shift the problem to our basic problem of identifying the best alternative routes (beside long running time). However, good alternatives are often a *k* shortest path with *k* being very large. Consider the following situation: From *s* to the *t* exist two long different highways, where the travel time on one highway is 5 minutes longer. To reach the highways we need to drive through the city. For the number of different paths through the city to the faster highway which travel time is not more than 5 minutes longer than the fastest path, we have a combinatorial explosion. The number of different paths is exponential in the number of nodes and edges in the city as we can independently combine short detours (around a block) within the city. That means, it is not feasible to compute all shortest paths until we discover the alternative path on the slightly longer highway. So we consider this method rather impractical for computing alternatives.

## 5.3 Disjoint Paths

Another approach are disjoint paths. We iteratively compute the shortest path, add it to the solution and delete all edges of the path in the base graph. That means, two alternatives are not allowed to share a single edge. This can produce very bad results. For example, if only one edge is incident to the start node, only one path is computed although there can exist meaningful alternatives. Also, if there exists a short subpath which is a bottle neck from start to destination, only one path can use this bottle neck and all others will include a large detour.

## 5.4 Plateau

The Plateau method [CAM09] identify fast highways (plateaus) and selects the best routes based on the full path length and the highway length. In more detail, we perform one regular Dijkstra [Dij59] from $s$ to all nodes and one backward Dijkstra from $t$ which uses all directed edges in the other direction. Then, we intersect the shortest path tree edges of both Dijkstra's. The resulting set consists of simple paths, we call each of those simple paths a *plateau*. We can complete a plateau to a *s-t*-path by a shortest path from $s$ to the simple path and a shortest path from the simple path to $t$. Here we can use the results of both Dijkstra's we computed before. Now we need to select the best alternative paths derived from the plateau. Therefore, we rank them by the length of the corresponding *s-t*-path and the length of the plateau, i.e. rank = (path length − plateau length). A plateau reaching from $s$ to $t$ would be 0, the best value. To ensure that the shortest path in the base graph is always the first path, we can prefer relaxed edges of the Dijkstra from $s$ during the backward Dijkstra of $t$ on a tie.

Plateau routes look good at first glance, although they may contain severe detours. In general, a plateau alternative can be described by a single via node, this is the biggest limitation of this method. Partitioning the shortest paths in subpaths and computing Plateau alternatives for each subpath could help to alleviate this problem.

## 5.5 Via Points of Bidirectional Speedup Techniques

The forward and backward search space of bidirectional speedup techniques, e.g. [SS06, SS07, GSSD08], usually meet at more than one node. Every meeting node represents an alternative route. The advantage of this method are the very fast computation, however the number and quality of the alternative routes is usually impossible to control. Flatter hierarchies with less levels usually have more meeting nodes and therefore provide potentially more and better alternative routes. This is interesting, as for shortest path computations, flatter hierarchies usually provide *less* speedup. Therefore, older techniques may have a revival for computing alternative routes.

## 5.6 Penalty

We propose a new approach to compute alternatives iteratively. Based on an evolving set of alternative routes, we adjust the weights in the graph so that a new shortest path computation will give a new (hopefully) meaningful route. The basic idea is that we compute the shortest path, add it to our solution, increase the edge weights on this path and start from the beginning until we are satisfied with our solution. The new shortest path is likely to be different from the last one, but not completely different, as some subpaths may still be shorter than a full detour (depending on the increase). The crucial point of this method is how we adjust the edge weights after each shortest path computation. We present an assortment of possibilities with which the combination results in meaningful alternatives.

First, we want to increase the edge weights of the last computed shortest path. We can add an absolute value on each edge of the shortest path, but this depends on the assembly and structure of the graph and penalizes short paths with many edges. We can by-pass this by multiplying the edge weights with a factor $k$, e.g. $k \in [1..2]$. The higher the factor (penalty), the more deviates the new shortest path from the last one.

Beside directly adding a computed shortest path to the solution, we can also first analyse the shortest path. If the shortest path provides us with a good alternative (e.g. is different and short enough), we add it to our solution. If not, we adjust the edge weights accordingly and recompute a shortest path.

Consider the case, the first part of the route has no meaningful alternative but the second part has 5. That means, that the first part of the route is likely to be increased several times during the iterations. In this case, we can get a shortest path with a very long detour on the first part of the route. To circumvent this problem, we can limit the number of increases of a single edge or just lower successive increases. If a new shortest path does not increase the weight of at least one edge, we are finished. This provides us with a natural saturation of the number of alternatives.

Another problem is, that the new shortest path can have many small detours (hops) along the route compared to the last path. Consider the following example: The last path is a long motorway and the new shortest path is *almost* equal to the last one, but at the middle of the motorway, it contains a very short detour (hops) from the long motorway on a less important road (due to the increase). There can occur many of those small hops, those look unpleasant for humans and contain no real alternative. In the alternative graph, this increases the number of decision edges while having no substantial positive effect on other attributes. To alleviate this problem, we propose several methods: First, we can not only increase the weights of edges on the path, but also of edges around the path (a tube). This avoids small hops, as edges on potential hops are increased and are therefore probably not shorter. The increase of the edges around the path should be decreasing with the distance to the path. Still, we penalize routes that are close to the shortest path, although there can be a long, meaningful alternative close to the shortest path. To avoid this, we can increase only the weights of the edges, which leave and join edges of the current alternative graph. We call this increase *rejoin penalty*. It should be additive and dependent on the general increase factor $k$ and the distance from $s$ to $t$, e.g. *rejoin penalty* $\in [0..(k-1.0) \cdot d(s,t)]$. This avoids small hops and reduces the number of decision edges in the alternative graph. The higher the *rejoin penalty* the less decision edges in the alternative graph. In some cases, we want more decision edges at the beginning or the end of the route, for example to find all spur routes to the highways. Therefore, we can grade the *rejoin penalty* according to the current position (*pos*, Section 3). Another possibility to get rid of small hops, is to allow them in the first place but remove them later in the alternative graph (Section 5.8).

To decrease the *coefficient of variation* (Section 3), we introduce another penalty: Before each computation of a shortest path, we compute the difference of the average number of edges to the actual number at edges at its position for every edge. If the number of edges is lower than the average, we add a penalty for this edge to lower the difference and therefore the coefficient of variation. This penalty should depend on the coefficient of variation.

A straightforward implementation of the Penalty method iteratively computes shortest paths using the Dijkstra algorithm. However, there are more sophisticated speedup techniques that can handle a reasonable number of increased edge weights [SS07]. Therefore we hope that we can efficiently implement the Penalty method.

## 5.7 Combinations

The Penalty method operates in general on a preexisting set of alternative routes and computes a new one. Therefore, a preprocessing based on any other method is possible. For example, the combination of the Plateau and Penalty method can produce an algorithm that is superior to a single one.

## 5.8 Refinements / Post Processing

The heuristics above often produce alternative graphs that can be easily improved by local refinements. For example alternatives that have most edges in common with other alternatives, should be removed as they provide no "real" alternative. The refinement helps to enforce the constraint on the number of decision edges.

# 6 Different Edge Weights

The methods to compute an alternative graph just depend on a single edge weight function (except Pareto). Therefore, we can use several different edge weight functions to independently compute alternative graphs. The different edge weights are potentially orthogonal to the alternatives and can greatly enhance the quality of our computed alternatives. When we combine the different alternative graphs in a single one, and want to compute its attributes of Section 3, we need to specify a main edge weight function, as the attributes also depend on the edge weights.

# 7 Conclusion

Our main contribution is a new way to characterize alternative routes that may look more natural to humans. Furthermore, we propose methods to compute such.